\documentclass[onecolumn,aps,showpacs]{revtex4}
\usepackage{graphicx,psfrag,dsfont}
\usepackage{slashed,amssymb}

\def\no{\noindent}
\newcommand{\nn}{\nonumber}

\begin{document}
 
\title{
Spontaneous mass generation due to phonons in a two-dimensional Dirac fermion system
}

\author{Andreas Sinner  and Klaus Ziegler}

\affiliation{Institut f\"ur Physik, Universit\"at Augsburg, D-86135 Augsburg, Germany}

\begin{abstract}
Fermions with one and two Dirac nodes are coupled to in-plane phonons to study a spontaneous transition 
into the Hall insulating phase. At sufficiently strong electron-phonon interaction a gap appears in the spectrum of fermions, 
signaling a transition into a phase with spontaneously broken parity and time-reversal symmetry. The structure of elementary excitations above the gap in 
the corresponding phase reveals the presence of scale invariant parity breaking 
terms which resemble Chern-Simons excitations. Evaluating the Kubo formula for both models we find quantized Hall 
plateaux in each case, with conductance of binodal model exactly twice as large as of the mononodal model.

\end{abstract}


\pacs{05.60.Gg, 72.10.Bg, 73.22.Pr}

\maketitle

\section{Introduction}
\label{sec:Intro}

\no
Because of their exceptional transport properties, two-dimensional electronic systems with Dirac-like 
low-energy spectrum of quasiparticles have been a subject of intense research for several 
decades~\cite{Fradkin1986,Lee1993,Ludwig1994,Zirnbauer1996,Ziegler1997,Gusynin2006}.
They provide an ideal platform for the simulation of quantum field theory models at low 
energies~\cite{Cirac2010,Korepin2013,Esslinger2014,Szameit2010,Szameit2011}.
As a model for Dirac-like systems in condensed matter, e.g. for graphene, it is important
to study the effect of disorder and electron-phonon interaction. It turned out that the
transport properties for a Fermi energy near the Dirac node are very robust against 
disorder~\cite{Morozov08,Chen08}.
This has also been supported by theoretical studies~\cite{Ludwig1994,Ziegler1997} and was recently attributed to
an attractive fixed point in the renormalization-group flow~\cite{CSPaper2016}. Lattice distortions in the
form of phonons are expected to have a more severe effect on transport because
they may open an electronic gap for sufficiently strong electron-phonon coupling
~\cite{Basko2008,Benfatto2009,Ziegler2011,Ziegler2011a,Politano15,CSPaper2016}. A gap opening
was also advocated by assuming a Kekul\'e distortion of the honeycomb lattice~\cite{Mudry2007,JakiwPi2007}. 

\vspace{1mm}
\no
A gap in the spectrum of a single Dirac node gives rise to quantized Hall 
current and to the Chern-Simons electrodynamics of the coupled gauge 
field~\cite{McCoy1979,Semenoff1984,Redlich1984,Fradkin1994,Dunne1999,Kondo1995,Froehlich2013}. 
On a bipartite lattices, though, nodes appear pairwise and the corresponding excitations differ by 
the chirality. A uniform Dirac mass with the same sign at each cone preserves both chiral and 
time-reversal symmetries. Then the Hall currents from the two nodes cancel each other, leading
to a vanishing Hall conductivity. To prevent
this cancellation we can add a periodic flux~\cite{Haldane1988} or a periodic
spin texture~\cite{Hill2011} to create different signs of the Dirac mass at the two nodes. Thus,
it is sufficient to study the Hall conductivity at a single massive Dirac fermion and to assume that
the other Dirac fermion appears with the opposite mass sign. This idea was employed to study the 
effect of phonons on the Hall conductivity in the case of weak electron-phonon 
interaction~\cite{CSPaper2016}. The effective Chern-Simons term gives the correct values
of the two Hall plateaux, but contributions from other quantum fluctuations generate small corrections. 
Moreover, the theory is unstable at sufficiently strong electron-phonon interaction. Thus, it would
be interesting to analyze the Hall conductivity in the regime of strong electron-phonon interaction
and to consider spontaneous Dirac mass generation.  
We will see that in this case the Chern-Simons field theory yields exactly two Hall plateaux.
For this purpose we address the issue of spontaneous chiral symmetry breaking and fermionic 
gap generation by monochromatic in-plane phonons, which play the role of a gauge field. 
While we suggested an onset of a new phase already in that older article, a 
detailed analysis of the phase transition and corresponding phase are presented below. 
For this we investigate a particular representation of the initial model which takes advantage of a 
Hubbard-Stratonovich decoupling in terms a matrix-valued quantum fields and account for all hidden
soft modes. This representation allows us not only to capture the fermionic mass generation, 
but also to determine the structure of the quantum 
fluctuations around the corresponding vacuum state to Gaussian order. We demonstrate that the
Gaussian term of the effective action exhibits a scale invariant term, 
which breaks the parity and time-reversal symmetry. Usually, these excitations are linked to a
Chern-Simons term. However, in the present case the topological term does not exhibit the 
standard form of a Chern-Simons term. The Hall conductivity calculated from 
the Kubo formula is non-zero for a single chiral Dirac fermion as well
as for a pair of two chiral fermions with different chiralities with the known universal values.

\section{Single node model}
\label{sec:EPModel2}

\no
We start our considerations with a model that describes a single copy of massless Dirac fermion 
coupling to a single in-plane phonon mode via spatial currents. For details of this coupling
we also refer to our earlier work Ref.~\cite{CSPaper2016}. The $2+1$-dimensional action of 
this model then reads 
\begin{eqnarray}
\label{eq:InAct}
{\cal S}[\bar\psi,\psi,A]
\label{eq:ElPh1}
&=& \frac{1}{2g} \vec A\cdot\vec A + \bar\psi\slashed\partial\psi + \frac{i}{\sqrt{2}} \vec A\cdot\vec j, 
\end{eqnarray}
where $\vec A = (A^{}_1,A^{}_2)$ is a two-component massive bosonic vector field which models the
dispersionless in-plane phonon with the mass $g^{-1}$. 
The slow lattice dynamics $\sim(\partial_\tau A_\mu)^2$ can be neglected in comparison to the much faster electron dynamics. 
In this approximation the phonon dynamics is generated by the fermion dynamics due to
the electron-phonon interaction. $\vec j = \bar\psi\vec\gamma\psi$ is the vector
consisting of two spatial fermionic currents $j^{}_{\mu=1,2} = \bar\psi\gamma^{}_{\mu}\psi$. 
The fermion kinetic term reads $\bar\psi\slashed\partial\psi\equiv \sum_{i=0,1,2}\bar\psi\partial^{}_i
\gamma^{}_i\psi$, $\bar\psi=\psi^\dag\gamma^{}_0$, $\psi$ denotes a two-component spinor (Dirac) fermion, 
and $\psi$ and $\bar\psi$ are independent integration variables. The set of $\gamma$-matrices is chosen
to be $\left\{\gamma^{}_0,\gamma^{}_1,\gamma^{}_2\right\}=\{\sigma^{}_3,-\sigma^{}_2,\sigma^{}_1\}$ 
with $\sigma^{}_i$ denoting the Pauli matrices in the usual representation, which fulfill the 
usual anticommutation relations
\begin{equation}
 \{ \gamma^{}_i,\gamma^{}_j\} = 2\delta^{}_{ij}.
\end{equation}
Below we will call the two-dimensional unity matrix the third gamma matrix, i.e. 
$\gamma^{}_3=\sigma^{}_0$, which commutes with all two-dimensional matrices. Therefore, any complex 
two-dimensional matrix $\Gamma$ can be uniquely expanded into the basis of $\gamma$-matrices
\begin{equation}
\label{eq:Decomp}
\Gamma = \sum_{\alpha=0,1,2,3}\Gamma^{\alpha}\gamma^{}_\alpha \equiv \Gamma^\alpha\gamma^{}_\alpha,
\end{equation}
with complex decomposition elements $\Gamma^\alpha$ and Einstein summation convection used in the second 
equality. 

\vspace{1mm}
\no
Integrating the phonon fields $A^{}_\mu$ generates an interaction term for the fermions $({g}/{4})(\bar\psi\gamma^{}_\mu\psi)^2$, which can be rewritten with the Fierz identity~\cite{Fierz1937,Thomaz1987}
as the repulsive Hubbard interaction $-\det(\sigma_\mu)(g/2)(\psi^\dag\psi)^2=(g/2)(\psi^\dag\psi)^2$.
This resembles the standard Thirring current--current interaction~\cite{Fradkin1994}, with one crucial difference, though, that we have only two spatial currents here.  The reader might be puzzled by the
repulsive interaction, because phonons usually lead to
an attractive electron-electron interaction causing a superconducting instability. This puzzle is due to the fact that in our case the phonons couple to the fermion currents~\cite{Stauber2007,amorim16}, 
whereas in the BCS theory they couple to the electron density.

\vspace{1mm}
\no
In the presence of phonons, the Hamiltonian reads
\begin{equation}
 H =  \nabla\cdot\gamma + \frac{i}{\sqrt 2}A\cdot\gamma, 
\end{equation}
where  $\nabla$ contains spatial derivatives only. The time reversal symmetry is defined as 
\begin{equation}
\label{eq:TRSc1}
\gamma^{}_2H^{\rm T}_{A\to-A}\gamma^{}_2 =  - H,
\end{equation}
where the operator T denotes the transposition on both spinor and position spaces and $A\to-A$  means the reflection of the phonon field. On another hand, the parity or chiral symmetry is given by
\begin{equation}
\label{eq:Chir1C1}
\gamma^{}_0H\gamma^{}_0 = -H. 
\end{equation}
Obviously, the appearance of the mass term which couples to the bilineal form $(\bar\psi\psi)$ breaks both symmetries.

\vspace{1mm}
\no
The action (\ref{eq:ElPh1}) is quadratic in both, fermions and bosons, and offers a number of 
possible ways for the treatment. We integrate out the bosons:
\begin{eqnarray}
\int\frac{{\cal D}\vec A}{\cal N} ~ \exp\left(\displaystyle -\frac{1}{2g} 
\vec A\cdot\vec A - \frac{i}{\sqrt{2}}\vec A\cdot\vec j \right)
&=& \exp\left(\displaystyle -\frac{g}{4} \vec j\cdot\vec j\right),
\end{eqnarray}
i.e. a repulsive interaction term between spatial currents appears in the action. It 
is possible to regroup it as
\begin{equation}
\label{eq:Int1}
\frac{g}{4} \vec j\cdot\vec j =
\frac{g}{4} \int d^3x ~ (\bar\psi\gamma^{}_\mu\psi)^2 = -\frac{g}{4}{\rm Tr}\int d^3x~(\gamma^{}_\mu 
\psi\bar\psi)(\gamma^{}_\mu \psi\bar\psi) = -\frac{g}{4}{\rm tr}{\cal D}^{2}_\mu,
\end{equation}
where the current density matrix ${\cal D}^{}_{\mu=1,2} = \gamma^{}_{\mu=1,2}\psi\bar\psi$, 
the summation over $\mu$ in Eq.~(\ref{eq:Int1}) is understood and the sign change is because 
$\psi$ is an anticommuting Grassmannian variable. In the chosen notation, the operator $\rm Tr$ 
denotes the trace on the Dirac space, while operator $\rm tr$ is the trace on both Dirac and 
position spaces. The interaction is quadratic in ${\cal D}^{}_\mu$ and can be decoupled by a 
suitable matrix Hubbard-Stratonovich transformation
\begin{eqnarray}
\label{eq:intmid1}
\exp\left(\displaystyle \frac{g}{4} {\rm tr}{\cal D}^{2}_\mu\right) & = & 
\int \frac{{\cal D}{\cal Q}}{\cal N}~
\exp\left(\displaystyle -\frac{1}{2g}{\rm tr}{\cal Q}^2_\mu - \frac{1}{\sqrt{2}} 
\bar\psi{\cal Q}^{}_\mu\gamma^{}_\mu \psi \right),\;\;\;
{\cal N} =  \int {{\cal D}{\cal Q}}~\exp\left(\displaystyle -\frac{1}{2g}{\rm tr}{\cal Q}^2_\mu\right),
\end{eqnarray}
with fields ${\cal Q}^{}_\mu$ that are 2$\times$2 complex matrices
\begin{equation}
{\cal Q}^{}_\mu = {\cal Q}^\alpha_\mu\gamma^{}_\alpha.
\end{equation}
These manipulations result in the following representation of our initial model 
\begin{equation}
{\cal S}[{\cal Q},\bar\psi,\psi] = \frac{1}{2g}{\rm tr}{\cal Q}^2_\mu 
+ \bar\psi\cdot\left[\slashed\partial + \frac{1}{\sqrt{2}}{\cal Q}^{}_\mu\gamma^{}_\mu\right]\psi
.
\end{equation}
In the functional integral the fermions can be integrated out to give as an effective action
\begin{equation}
\label{eq:LogAct1}
 {\cal S}[{\cal Q}] = \frac{1}{2g}{\rm tr}{\cal Q}^2_\mu - {\rm tr}\log\left[\slashed\partial 
 + \frac{1}{\sqrt{2}}{\cal Q}^{}_\mu\gamma^{}_\mu\right]. 
\end{equation}

\subsection{Kubo formula in matrix field representation}

\no
The field theoretical version of the Kubo formula we will work with reads in fermionic representation 
\begin{eqnarray}
\bar\sigma^{}_{\mu\nu} = \frac{2\pi}{\omega} \int d^3x \left[e^{-i\omega(x^{}_0-y^{}_0)} 
-1\right]\langle j^{}_{x,\mu}j^{}_{y,\nu} \rangle^{}_{\bar\psi\psi,A}, 
\end{eqnarray}
where $\langle \cdots \rangle^{}_{\bar\psi\psi,A}$ denotes the normalized functional integration
with respect to the action in Eq. (\ref{eq:ElPh1}). We can reorganize the local current densities as 
$\displaystyle j^{}_{x,\mu} =  (\bar\psi\gamma^{}_\mu\psi)^{}_x=-{\rm Tr}
\left(\gamma^{}_\mu\psi\bar\psi\right)^{}_x =-{\rm Tr}{\cal D}^{}_{x,\mu},$ and rewrite the Kubo
formula as
\begin{equation}
\bar\sigma^{}_{\mu\nu} = \frac{2\pi}{\omega} \int d^3x \left[e^{-i\omega(x^{}_0-y^{}_0)} -1\right] 
\langle {\rm Tr}{\cal D}^{}_{x,\mu}{\rm Tr}{\cal D}^{}_{y,\nu} \rangle^{}_{\bar\psi\psi,A} 
.
\end{equation}
This form is suitable to write the Kubo formula in terms of a correlation of the $\cal Q$-field.
The details of this transformation are given in Appendix~\ref{app:Kubo}. The trace is performed
by using the expansion Eq.~(\ref{eq:Decomp}) and by diagonalizing the matrix element in momentum space.
Thus, we obtain for the Kubo formula
\begin{eqnarray}
\label{eq:KuboApp}
\bar\sigma^{}_{\mu\nu} &=& \frac{16\pi}{\omega g^2}\int\frac{d^2q}{(2\pi)^2}\delta(\vec q)
\int\frac{dq^{}_0}{2\pi}~ \left[\delta(q^{}_0+\omega) - \delta(q^{}_0)\right]  
\langle{\cal Q}^{3}_{Q,\mu}{\cal Q}^{3}_{-Q,\nu}\rangle^{}_{{\cal Q}}.
\end{eqnarray}
All terms associated with the coordinate $y$ disappear due to translational invariance. 
We recognize further that it is sufficient to determine the inverse propagator matrix in the action 
to leading order in the frequency $\omega$ only.
To compute the conductivity we need to determine the explicit form of the non-linear 
action Eq.~(\ref{eq:LogAct1}) up to Gaussian (second) order.

\subsection{Saddle point analysis}

\no
Variation  of Eq.~(\ref{eq:LogAct1}) with respect to the field $\cal Q$ yields the saddle--point equation
\begin{equation}
0 = \delta{\cal S}  = {\rm tr}~\delta{\cal Q}^{}_\alpha \left\{\frac{1}{g}{\cal Q}^{}_\alpha 
- \frac{1}{\sqrt{2}} \gamma^{}_\alpha \left[\slashed\partial +\frac{1}{\sqrt{2}}{\cal Q}^{}_\mu
\gamma^{}_\mu \right]^{-1}\right\}.
\end{equation}
For translationally invariant solutions $\bar{\cal Q}$ the expression in the curly brackets 
is diagonal in position space
\begin{equation}
\label{eq:Saddle1}
\bar{\cal Q}^{}_\alpha = \frac{g}{\sqrt{2}}\gamma^{}_\alpha \left[\slashed\partial 
+\frac{1}{\sqrt{2}}\bar{\cal Q}^{}_\mu\gamma^{}_\mu \right]^{-1}_{rr},
\end{equation}
where $\alpha=1,2$ is fixed and not summed over. The solution is found with the ansatz 
$\displaystyle \bar{\cal Q}^{}_\alpha = \frac{m}{\sqrt{2}}\gamma^{}_\alpha$, with a free 
sign of $m$, which yields
\begin{equation}
m = g  \left[\slashed\partial + m\right]^{-1}_{rr} = g \int\frac{d^3Q}{(2\pi)^3} 
~ [-i\slashed Q + m ]^{-1}.
\end{equation}
Besides the trivial solution $m=0$ we also detect a non-trivial solution giving a non-zero value of 
$m$ which 
describes a mass generation of the fermions and thus breaks the time reversal symmetry. 
Dropping rotationally non-invariant contributions we get 
\begin{equation}
\label{eq:Saddle2}
1 = \int\frac{d^3Q}{(2\pi)^3} ~\frac{g}{Q^2+m^2},
\end{equation}
where $Q^2=q^2_0+q^2$, $q^{}_0$ denotes the zero-temperature Matsubara frequency.
At zero temperature there is an infinite integral with respect to $q_0$:
\begin{eqnarray}
\frac{1}{g} &=& \intop^{\infty}_{-\infty}\frac{dq^{}_0}{2\pi}\intop^{\Lambda}
\frac{d^2\vec q}{(2\pi)^2}~\frac{1}{q^2_0+q^2+m^2} = \frac{1}{4\pi}[\sqrt{\Lambda^2+m^2} - |m|]
,
\end{eqnarray}
which leads to
\begin{equation}
|m| = \frac{g}{8\pi}\left[\Lambda^2 - \left(\frac{4\pi}{g}\right)^2\right] 
\sim \frac{g}{\pi}\left(\frac{2\pi}{g^{}_c}\right)^2\Theta(g-g^{}_c) + {\cal O}(g^2),
\end{equation}
where $g^{}_c={4\pi}/{\Lambda}$ is the critical value of the coupling strength, 
above which the fermionic gap is non-zero.

\subsection{Gaussian fluctuation term of the action in the gapped phase}

\no
Shifting matrix fields in Eq.~(\ref{eq:LogAct1}) with respect to the saddle point
\begin{equation}
{\cal Q}^{}_\alpha \to {\cal Q}^{}_\alpha + \frac{m}{\sqrt{2}}\gamma^{}_\alpha, 
\end{equation}
and expanding the logarithm in powers of the inverse mass using $\log[1+x]\sim x-{x^2}/{2}$,
the expansion of the action up to second order reads
\begin{eqnarray}
{\cal S}^{}_G[{\cal Q}] &=& \frac{1}{2g}{\rm tr}{\cal Q}^2_\mu 
+ \frac{1}{4}{\rm tr}[G{\cal Q}^{}_\mu\gamma^{}_\mu]^2 
={\rm tr} {\cal Q}^a_\mu{\cal Q}^b_\nu \left\{\frac{1}{2g}\gamma^{}_a\gamma^{}_b 
\delta^{}_{\mu\nu} + \frac{1}{4}G\gamma^{}_a\gamma^{}_\mu G \gamma^{}_b \gamma^{}_\nu\right\}
\end{eqnarray}
with $G^{-1}=\slashed\partial+ m$. 
This  means that the kernel of the quadratic form is an 8$\times$8-matrix. For practical 
calculations it is useful to replace $1/g$ using Eq.~(\ref{eq:Saddle2}).
Assembling the field components as a vector 
\begin{equation}
\vec{\cal Q} = ( {\cal Q}^{0}_{1} , {\cal Q}^{1}_{1}, {\cal Q}^{2}_{1},  {\cal Q}^{3}_{1},    
{\cal Q}^{0}_{2},  {\cal Q}^{1}_{2},     {\cal Q}^{2}_{2},   {\cal Q}^{3}_{2}   )^{T}, 
\end{equation}
we get for the Gaussian term in Fourier representation
\begin{equation}
{\cal S}^{}_G[{\cal Q}] = \int\frac{d^3P}{(2\pi)^2}~\vec {\cal Q}^{}_P {\bf \Pi}^{-1}(P)
\vec {\cal Q}^{}_{-P},
\end{equation}
with the matrix element of the inverse propagator matrix ${\bf \Pi}^{-1}_{a\mu;b\nu}(P)$ given with 
the help of Feynman-parametrization as
\begin{equation}
{\bf \Pi}^{-1}_{a\mu;b\nu}(P) = \int\frac{d^3Q}{(2\pi)^3}~\frac{\delta^{}_{ab}
\delta^{}_{\mu\nu}}{Q^2+m^2} + \frac{1}{4}\int_0^1dx\int\frac{d^3Q}{(2\pi)^3}~
\frac{{\rm Tr}\{[i(\slashed Q-(1-x)\slashed P)-m]\gamma^{}_a
\gamma^{}_\mu[i(\slashed Q+x\slashed P)-m]\gamma^{}_b\gamma^{}_\nu\}}
{[Q^2+m^2+x(1-x)P^2]^2}.
\end{equation}
Up to quadratic order in the gradient expansion the inverse matrix propagator can be written 
in a block form
\begin{equation}
\label{eq:InvProp}
{\bf \Pi}^{-1}(P) = \left( 
\begin{array}{cc}
 {\bf A} & {\bf B} \\ 
 {\bf C} & {\bf D}
\end{array}
\right).
\end{equation}
The explicit expressions for every $4\times 4$ block are given in Appendix~\ref{app:sinmod}. 
Here we notice with the help of the saddle-point condition Eq.~(\ref{eq:Saddle2}) that the following 
relations between momentum independent terms of the gradient expansion shown in
Eq.~(\ref{eq:app:mas}) hold
\begin{equation}
\label{eq:MassRel} 
M^{}_1 + \Delta^{}_1 = \frac{1}{g},\;\;\; M^{}_4 - \Delta^{}_1 = \frac{1}{g}. 
\end{equation}
The inverse propagator ${\bf \Pi}^{-1}(P)$ 
is invariant under matrix transposition and momentum mirroring, which is a finger print of the 
broken time reversal invariance:
\begin{equation}
{\bf \Pi}^{-1} (P) = [{\bf\Pi}^{-1}(-P)]^{\rm T}. 
\end{equation}
All phonon modes are massive with positive masses (cf. Appendix~\ref{app:sinmod}); i.e.,
the existence of the bosonic functional integral is secured. 

\vspace{2mm}
\no
For computing the conductivity we can put all spatial momenta in the inverse propagator to zero. 
The $8\times8$ matrix then reads
\begin{equation}
 {\bf \Pi}^{-1} (p^{}_0) \sim  \left(
\begin{array}{cccccccc} 
M^{}_1 &  0 & 0 & \frac{isp^{}_0}{16\pi} & \frac{sp^{}_0}{16\pi} & 0 & 0 & i \Delta^{}_1 \\
\\
0 & M^{}_2 & 0 & 0 & 0 & 0 & \Delta^{}_2 & 0 \\
\\
0 & 0 & M^{}_3 & 0 & 0 & \Delta^{}_3 & 0 & 0 \\
\\
-\frac{isp^{}_0}{16\pi} & 0 & 0 & M^{}_4 & -i\Delta^{}_1 & 0 & 0 & -\frac{sp^{}_0}{16\pi} \\
\\
-\frac{sp^{}_0}{16\pi} & 0 & 0 & -i\Delta^{}_1 & M^{}_1  & 0 & 0 & \frac{isp^{}_0}{16\pi} \\
\\
0 & 0 & \Delta^{}_3 & 0 & 0 & M^{}_3 & 0 & 0 \\
\\
0 & \Delta^{}_2 & 0 & 0 & 0 & 0 & M^{}_2 & 0 \\
\\
i\Delta^{}_1 & 0 & 0 & \frac{sp^{}_0}{16\pi} & -\frac{isp^{}_0}{16\pi} & 0 & 0 & M^{}_4 
\end{array}
\right),
\end{equation}
which can be inverted analytically. The time-reversal symmetry breaking term is referred to as the 
Chern-Simons term, acquires the following form 
\begin{equation}
\label{eq:CSt1} 
{\cal S}^{1C}_{CS} = \frac{{\rm sgn}(m)}{16\pi}\int\frac{d^3P}{(2\pi)^3}~p^{}_{0}\left[(-1)^a
~\epsilon^{}_{\mu\nu} {\cal Q}^a_{\mu, P}{\cal Q}^a_{\nu, -P} + i\epsilon^{ab} {\cal Q}^a_{\mu, P}{\cal Q}^b_{\mu,-P} \right], 
\end{equation}
where we replaced $s={\rm sgn}(m)$. The upper indices run over $a,b=0,3$ and the corresponding
antisymmetric tensor has the components $\epsilon^{ab}=+1$ for $a=0,b=3$ and $\epsilon^{ab}=-1$ 
for $a=3,b=0$, while the lower indices run over $\mu,\nu=1,2$ with  the conventional antisymmetric 
tensor $\epsilon^{}_{\mu\nu}$. The spatial components of the Chern-Simons tensor cannot be cast into 
a similarly simple expression.

\subsection{Hall conductivity for a single node}

\no
In order to calculate dc conductivities from the Kubo formula Eq.~(\ref{eq:KuboApp}) we fix lower field
indices to $\mu=1,\;\nu=1$ for longitudinal conductivity and $\mu=1,\;\nu=2$ for the Hall conductivity.
Hence we get for the longitudinal conductivity 
\begin{equation}
\langle{\cal Q}^{3}_{P,1}{\cal Q}^{3}_{-P,1}\rangle^{}_{{\cal Q}} = \frac{1}{2}{\bf \Pi}^{}_{44}(-P),
\end{equation} 
where the factor $1/2$ appears from the Gauss integral, which obviously gives a zero conductance. 
For the Hall conductivity we obtain
\begin{equation}
\langle{\cal Q}^{3}_{p^{}_0,1}{\cal Q}^{3}_{-p^{}_0,2}\rangle^{}_{{\cal Q}} = \frac{1}{2}
{\bf \Pi}^{}_{48}(-p^{}_0) \sim-\frac{sp^{}_0}{32\pi}\frac{(M^{}_1+\Delta^{}_1)^2}{(M^{}_1M^{}_4+\Delta^{2}_1)^2},
\end{equation} 
where the $p^{}_0$-dependence in the denominator is omitted. Putting the matrix element into the 
Kubo formula Eq.~(\ref{eq:KuboApp}) and using mass relations Eq.~(\ref{eq:MassRel}) we finally get
to Gaussian order the well known result for the Hall conductivity of a single Dirac cone:
\begin{equation}
 \bar\sigma^{\rm dc}_{12} \sim\frac{1}{2} {\rm sgn}(m).
\end{equation}
This is the first term in Eq.~(\ref{eq:CSt1}) which  determines the Hall conductivity and
has more similarity to the conventional Chern-Simons term.

\section{Binodal model}

\no
Next we shall demonstrate that our analysis is also applicable to graphene-like systems
with two massless Dirac nodes. Our intention is to study spontaneous 
generation of two Dirac masses with opposite signs through a sufficiently
strong electron-phonon interaction. For this purpose we start with the Hamiltonian
\begin{equation}
\label{eq:H2C}
{\cal H} = \Psi^\dag\cdot\left(
\begin{array}{cccc}
0  &  i \partial^{}_1 + \partial^{}_2  &  A^{}_1 - i A^{}_2  &  0 \\
\\
i\partial^{}_1 - \partial^{}_2 & 0 & 0 & - A^{}_1 - i A^{}_2 \\
\\
A^{}_1 + i A^{}_2 & 0 & 0 & i\partial^{}_1 - \partial^{}_2 \\
\\
0 & -A^{}_1 + i A^{}_2 & i\partial^{}_1 + \partial^{}_2 & 0
\end{array}
\right)\Psi,
\end{equation}
where $\Psi=(\Psi_{11},\Psi^{}_{12},\Psi^{}_{21},\Psi^{}_{22})^{\rm T}$ is the 4-component Dirac 
bispinor and the phonon field couples the two Dirac nodes. 
This Hamiltonian is obviously invariant with respect to the time-reversal and
parity transformations: interchanging  both fermionic copies and transposing them on position 
space maps the Hamiltonian onto itself. We introduce the following notation for 4$\times$4 matrices: 
\begin{equation}
\Sigma^{}_{ij} = \sigma^{}_{i}\otimes\sigma^{}_{j},\;\;\; i,j=0,1,2,3,
\end{equation}
with $\sigma^{}_i$ denoting the Pauli matrices in the usual representation. These are 
15 traceless matrices which represent a particular representation of the group of $SU(4)$-transformations
and a 4-dimensional unit matrix $\Sigma^{}_{00}$. The analogon of decomposition Eq.~(\ref{eq:Decomp}) for 
any complex $4\times4$-matrix reads $\Gamma=\Gamma^{ij}\Sigma^{}_{ij}$.
In analogy to Eq. (\ref{eq:ElPh1}) we introduce the action
\begin{equation}
{\cal S}[A,\psi^\dag,\psi] = \frac{1}{2g}\vec A\cdot\vec A + \psi^\dag\cdot\left[\partial^{}_\tau
\Sigma^{}_{00} -i\vec\partial\cdot\vec j + \vec A\cdot\vec \Pi \right]\psi,
\end{equation}
where we introduce the current vector $\vec j = \{\Sigma^{}_{01},\Sigma^{}_{32}\} $ and the 
coupling vector $\vec\Pi=\{\Sigma^{}_{13},\Sigma^{}_{20}\}$. The choice of the phonon part is 
dictated by the properties of the $C^{}_6$ group: According to the notation developed in 
Ref.~\cite{Basko2008}, electronic and phononic sectors on the space of 4$\times$4 matrices are 
defined by two mutually commuting subsets of the full set of 16 generators of the SU(4) group, 
each consisting of three matrices obeying Pauli anticommutation relations. The choice of 
those subsets is not unique, though. In this paper we use the representation with $\{\Sigma^{}_{01},
\Sigma^{}_{32},\Sigma^{}_{33}\}$ for the set, which emphasizes the different chiralities of the
fermions within a Weyl pair. The second set is chosen to be $\{\Sigma^{}_{30},
\Sigma^{}_{11},\Sigma^{}_{21}\}$, such that the phonons appear in the off-diagonal blocks of the
Hamiltonian. The corresponding coupling matrices appear as $\vec\Pi=\{-\Sigma^{}_{21}\Sigma^{}_{32},
\Sigma^{}_{21}\Sigma^{}_{01}\}$ for an appropriate definition of the phonon field components. 
Alternatively, one can also choose $\left\{\Sigma^{}_{01},\Sigma^{}_{02},\Sigma^{}_{03}\right\}$ 
and $\{\Sigma^{}_{20},\Sigma^{}_{30},\Sigma^{}_{10}\}$ which formally deals with a pair of Dirac 
fermions with the same chirality~\cite{Ziegler2011}. The phonons appear in the 
off-diagonal blocs again and couple to the matrices $\{\Sigma^{}_{11},\Sigma^{}_{12}\}$.
It will be shown below that both representations are equivalent at least to mean-field approximation.

\vspace{1mm}
\no

In analogy to the one cone model it is easy to identify the extended time-reversal and chiral symmetry, which will be spontaneously broken. 
In place of Eqs.~(\ref{eq:TRSc1}) and  (\ref{eq:Chir1C1})  we get here with the Hamiltonian Eq.~(\ref{eq:H2C})
\begin{equation}
H = -i\partial^{}_1\Sigma^{}_{01} - i\partial^{}_2\Sigma^{}_{32} + A^{}_1\Sigma^{}_{13} + A^{}_2 \Sigma^{}_{20},
\end{equation}
the similar symmetry breaking patterns for the time-reversal symmetry
\begin{equation}
\Sigma^{}_{02}H^{\rm T}_{A\to-A} \Sigma^{}_{02} = H,\;\;\; {\rm while }\;\;\; \Sigma^{}_{02} [H^{\rm T}_{A\to-A} + m\Sigma^{}_{33}]\Sigma^{}_{02} = H - m\Sigma^{}_{33},
\end{equation}
the transposition is to be made on all spaces, and for the chiral symmetry:
\begin{equation}
\label{eq:2Cchiral}
\Sigma^{}_{33} H\Sigma^{}_{33} = - H,\;\;\; {\rm while }\;\;\;\Sigma^{}_{33}[H + m\Sigma^{}_{33}]\Sigma^{}_{33} = - H + m\Sigma^{}_{33},
\end{equation}
that is, the parity between the nodal points, i.e. ultimately the sublattice symmetry of hexagonal lattice is no longer maintained.

\vspace{2mm}
\no
Following the procedure developed in the previous section, we integrate out the bosonic field 
$\vec A$, which creates a four-fermion interaction term. The latter can be rewritten as trace over 
the product of two matrices and can be decoupled by 4$\times$4 matrix fields $\cal Q$
\begin{equation}
\frac{g}{2} \left(\psi^\dag\Pi^{}_\mu\psi\right)^2 = -\frac{g}{2}{\rm tr}
\left(\Pi^{}_\mu\psi\psi^\dag\right)^2 \rightarrow \frac{1}{2g}{\rm tr}{\cal Q}^2_\mu 
+i\psi^\dag{\cal Q}^{}_\mu\Pi^{}_\mu\psi,
\end{equation}
where the decoupling field decomposes as ${\cal Q}^{}_\mu={\cal Q}^{ij}_\mu\Sigma^{}_{ij}$. 
The full action in this representation reads
\begin{equation}
{\cal S}[{\cal Q},\psi^\dag,\psi] = \frac{1}{2g}{\rm tr}{\cal Q}^2_\mu 
+\psi^\dag\cdot\left[G^{-1}_0 + i{\cal Q}^{}_\mu\Pi^{}_\mu\right]\psi,\;\;\; 
G^{-1}_0 = \partial^{}_\tau\Sigma^{}_{00} + i\vec \partial\cdot\vec j
 .
\end{equation}
Now we can integrate out the fermions to arrive at the bosonic action
\begin{equation}
\label{eq:BosAct}
{\cal S}[{\cal Q}] = \frac{1}{2g}{\rm tr}{\cal Q}^2_\mu - {\rm tr}\log\left[G^{-1}_0 
+ i{\cal Q}^{}_\mu\Pi^{}_\mu \right].
\end{equation}

\subsection{Saddle-point analysis}

\no
The variation of the action Eq.~(\ref{eq:BosAct}) gives the following saddle-point equations:
\begin{equation}
\label{eq:SPE}
\Pi^{}_\alpha\bar{\cal Q}^{}_\alpha = ig\left[G^{-1}_0 + i\bar{\cal Q}^{}_\mu\Pi^{}_\mu \right]^{-1}_{rr},
\end{equation}
where $\alpha$ is fixed and not summed over. We are interested in the solution that describes a 
spontaneous appearance of a mass which has different signs at different nodes. This requires 
a saddle point solution proportional to the matrix $\Sigma^{}_{33}$. The corresponding field 
configurations should be $\Pi^{}_\alpha\bar{\cal Q}^{}_\alpha \sim \Sigma^{}_{33}$; i.e.,
they differ for different $\alpha$:
\begin{eqnarray}
 \alpha = 1: & \displaystyle 
 m^{}_1 \Sigma^{}_{13}\Sigma^{}_{20}  = ig\int\frac{d^3Q}{(2\pi)^3}~\left[G^{-1}_0 
 +i \left(m^{}_1 \Sigma^{}_{20}\Sigma^{}_{13} + m^{}_2 \Sigma^{}_{13} \Sigma^{}_{20}\right)\right]^{-1}, \\
 \alpha = 2: & \displaystyle 
 m^{}_2 \Sigma^{}_{20}\Sigma^{}_{13} = ig\int\frac{d^3Q}{(2\pi)^3}~\left[G^{-1}_0 
 +i \left(m^{}_1 \Sigma^{}_{20}\Sigma^{}_{13} + m^{}_2 \Sigma^{}_{13} \Sigma^{}_{20}\right)\right]^{-1}.
\end{eqnarray}
Since matrices in each pair anticommute, their order matters:
\begin{eqnarray}
\alpha=1: & \displaystyle 
im^{}_1 \Sigma^{}_{33} = ig\int\frac{d^3Q}{(2\pi)^3}~\left[G^{-1}_0 + (m^{}_1 - m^{}_2)
\Sigma^{}_{33} \right]^{-1} \\
\alpha=2: & \displaystyle 
-im^{}_2 \Sigma^{}_{33} = ig\int\frac{d^3Q}{(2\pi)^3}~\left[G^{-1}_0 + (m^{}_1 
- m^{}_2)\Sigma^{}_{33} \right]^{-1}. 
\end{eqnarray}
Subtracting both sides of the equations yields
\begin{equation}
m^{}_1+m^{}_2 = 0,\;\; \curvearrowright \;\; m^{}_2= - m^{}_1,\;\; \curvearrowright m^{}_1-m^{}_2 
= 2m^{}_1 = m.
\end{equation}
Then, each of two equations reduces to 
\begin{equation}
\label{eq:SadP2}
m\Sigma^{}_{33} = 2g\int\frac{d^3Q}{(2\pi)^3}~\left[G^{-1}_0 + m\Sigma^{}_{33} \right]^{-1}. 
\end{equation}
To invert the matrix under the integral, we notice that the matrix $\Sigma^{}_{33}$ 
anticommutes with both currents. The inverse matrix then reads
\begin{equation}
[G^{-1}_0 +m\sigma^{}_{33}]^{-1} = [iq^{}_0\Sigma^{}_{00}+\vec q\cdot\vec j + m\Sigma^{}_{33}]^{-1} 
= \frac{-iq^{}_0\Sigma^{}_{00} + \vec q\cdot\vec j+m\Sigma^{}_{33}}{q^2_0+q^2+m^2},
\end{equation}
which yields, upon inserting it into Eq.~(\ref{eq:SadP2}), the non-trivial gap equation
\begin{equation}
\label{eq:BiNSP}
\frac{1}{g} = \int\frac{d^3Q}{(2\pi)^3} \frac{2}{Q^2+m^2} .
\end{equation}
This reproduces Eq.~(\ref{eq:Saddle2}), up to the factor of 2. Thus, all subsequent discussions
of the previous section apply here as well. 
Moreover, the saddle-point equation Eq.~(\ref{eq:BiNSP}) does not change if we change to the
representation of the electron-phonon coupling discussed in Ref.~\cite{Ziegler2011}, 
as is shown in Appendix~\ref{app:SPEmonochir}. 
Therefore, the field shift is individual for every matrix field
\begin{equation}
{\cal Q}^{}_\mu \to {\cal Q}^{}_\mu - (-1)^\mu\frac{m}{2}\Sigma^{(m)}_{\mu}, 
\end{equation}
where $\Sigma^{(m)}_1=\Sigma^{}_{20}$ and $\Sigma^{(m)}_2=\Sigma^{}_{13}$. 

\vspace{2mm}
\no
Eqs.~(\ref{eq:SadP2}) and (\ref{eq:BiNSP}) deserve a more extended discussions. In contrast 
to the case of a single cone, where the gap generation by phonons is not really surprising, taken 
the analogy to conventional superconductivity, the binodal case is more peculiar, since the gap 
generation accompanies the spontaneous breaking of the time-reversal and parity symmetry of the Hamiltonian. 
Since there should not be any gapless modes, the emerging phase would describe an insulator. 
However, because ${\rm Tr}[\Sigma^{}_{01}\Sigma^{}_{32}\Sigma^{}_{33}]\neq 0$, there is 
a Chern-Simons action and, consequently, a quantized Hall current. Hence, Eqs.~(\ref{eq:SadP2}) and 
(\ref{eq:BiNSP}) describe the transition of the system into the Chern insulating phase which 
takes place at sufficiently large values of the electron-phonon coupling.

\subsection{Gaussian fluctuations of the binodal model} 

\no
The Gaussian action in the present case is given by 
\begin{equation}
\label{eq:Gauss1}
{\cal S}^{}_G[{\cal Q}] = {\rm tr}\left\{ \frac{1}{2g}{\cal Q}^2_\mu - 
\frac{1}{2}G{\cal Q}^{}_\alpha\Pi^{}_\alpha G{\cal Q}^{}_\beta\Pi^{}_\beta\right\}, \;\;\; G^{-1} 
= G^{-1}_0 + m\Sigma^{}_{33}, 
\end{equation}
where the minus sign is because of the $i$ in front of $\cal Q$ in Eq.~(\ref{eq:BosAct}). 
Since fields $\cal Q$ are 4$\times$4 matrices, the kernel, (i.e. inverse bosonic propagator 
matrix) of the quadratic form of Gaussian action is a 32$\times$32 matrix. In this representation,
 Eq.~(\ref{eq:Gauss1}) can be written as:
\begin{equation}
\label{eq:Gauss2}
{\cal S}^{}_G[{\cal Q}] = 2\int\frac{d^3P}{(2\pi)^3}~
\left( 
\begin{array}{c}
\vec{\cal Q}^{}_1\\
\vec{\cal Q}^{}_2
\end{array}
\right)^{\rm T}_P
\left( 
\begin{array}{cc}
 {\bf A} & {\bf B} \\
 {\bf C} & {\bf D}
\end{array}
\right)
\left( 
\begin{array}{c}
\vec{\cal Q}^{}_1\\
\vec{\cal Q}^{}_2
\end{array}
\right)^{}_{-P}
\end{equation}
with the vector field
\begin{equation}
\vec{\cal Q}^{}_{i} = \left({\cal Q}^{00}_i,{\cal Q}^{01}_i,{\cal Q}^{02}_i,{\cal Q}^{03}_i, 
{\cal Q}^{10}_i,{\cal Q}^{11}_i,{\cal Q}^{12}_i,{\cal Q}^{13}_i,{\cal Q}^{20}_i,{\cal Q}^{21}_i,
{\cal Q}^{22}_i,{\cal Q}^{23}_i,{\cal Q}^{30}_i,{\cal Q}^{31}_i,{\cal Q}^{32}_i,{\cal Q}^{33}_i\right), 
\;\; i=1,2;
\end{equation}
i.e., each matrix block itself is a 16$\times$16 matrix. Here we only write the time-component of 
the Chern-Simons tensor:
\begin{equation}
\label{eq:CS2} 
{\cal S}^{2C}_{CS} = \frac{{\rm sgn}(m)}{8\pi}\int\frac{d^3P}{(2\pi)^3}~p^{}_0 
\left[i\epsilon^{ab}\left({\cal Q}^{ab}_{\mu,P}{\cal Q}^{ba}_{\mu,-P}-{\cal Q}^{aa}_{\mu,P}{\cal Q}^{bb}_{\mu,-P}\right) 
- (-1)^a\epsilon^{}_{\mu\nu}{\cal Q}^{ab}_{\mu,P}{\cal Q}^{ab}_{\nu,-P}\right],
\end{equation}
where the numbers run over $a,b=\{0,3\},\{1,2\}$ and ${\mu,\nu}=\{1,2\}$. The antisymmetric 
tensor with upper indices has following elements: $\epsilon^{03}=-\epsilon^{30}=\epsilon^{12}
=-\epsilon^{21}=1$. All matrix elements are given explicitly in Appendix~\ref{app:binmod}. 
Here we give a few of important relations for some of momentum independent parts form 
Eq.~(\ref{eq:app:2cmas})
\begin{equation}
\label{eq:MassAlg2}
M^{}_1 - \Delta^{}_1 = \frac{1}{g}, \;\;\; M^{}_5 + \Delta^{}_{1} = \frac{1}{g},\,\;\; 
M^{}_1 + M^{}_5 = \frac{2}{g}.
\end{equation}
As in the case of the single cone model, the inverse propagator matrix is 
invariant under simultaneous matrix transposition and momentum mirroring.

\subsection{Hall conductivity for two nodes}

\no
Proceeding in analogy to the one-cone model, cf. Appendix~\ref{app:Kubo} Eq.~(\ref{eq:app:current}), we can bring the Kubo formula into the 
$\cal Q$-matrix representation: 
\begin{equation}
\bar\sigma^{}_{\mu\nu} = -\frac{\pi}{2\omega g^2}\int d^3x 
~ \left[e^{-i\omega(x^{}_0-y^{}_0)}-1\right]\langle {\rm Tr}\left(j^{}_\mu\Pi^{}_a{\cal Q}^{}_a\right)^{}_x 
{\rm Tr}\left(j^{}_\nu\Pi^{}_b{\cal Q}^{}_b\right)^{}_y \rangle^{}_{\cal Q}
\end{equation}
Fixing $\mu=1$ and $\nu=2$, each trace leaves only two terms:
\begin{eqnarray}
 {\rm Tr}\left(j^{}_\mu\Pi^{}_a{\cal Q}^{}_a\right) & 
 = & {\rm Tr}\left({\cal Q}^{12}_{1}\Sigma^{}_{01}\Sigma^{}_{13}\Sigma^{}_{12} 
 + {\cal Q}^{21}_{2}\Sigma^{}_{01}\Sigma^{}_{20}\Sigma^{}_{21}\right) = 
 4(-i{\cal Q}^{12}_{1} + {\cal Q}^{21}_{2}),   \\
 {\rm Tr}\left(j^{}_\nu\Pi^{}_b{\cal Q}^{}_b\right) & 
 = & {\rm Tr}\left({\cal Q}^{21}_{1}\Sigma^{}_{32}\Sigma^{}_{13}\Sigma^{}_{21} 
 + {\cal Q}^{12}_{2}\Sigma^{}_{32}\Sigma^{}_{20}\Sigma^{}_{12}\right) = 
 4(-{\cal Q}^{21}_{1} -i {\cal Q}^{12}_{2}), 
\end{eqnarray}
which gives for the entire matrix element
\begin{equation}
{\rm Tr}\left(j^{}_\mu\Pi^{}_a{\cal Q}^{}_a\right)^{}_x {\rm Tr}\left(j^{}_\nu\Pi^{}_b{\cal Q}^{}_b\right)^{}_y = 
16\left(
i{\cal Q}^{12}_{1x}{\cal Q}^{21}_{1y} - i{\cal Q}^{21}_{2x}{\cal Q}^{12}_{2y} - {\cal Q}^{12}_{1x}{\cal Q}^{12}_{2y} - {\cal Q}^{21}_{2x}{\cal Q}^{21}_{1y}
\right),
\end{equation}
where we can immediately make the link to Eq.~(\ref{eq:CS2}) in order to recognize the sectors in the action which are responsible for the
Hall conductivity.  Performing the functional integration we obtain
\begin{eqnarray}
\bar\sigma^{}_{12} = \frac{8\pi}{\omega g^2}\int\frac{dq^{}_0}{2\pi}[\delta(q^{}_0+\omega)-\delta(q^{}_0)]\frac{1}{4}
\left\{{\bf \Pi}^{}_{26,10}(-q^{}_0) + {\bf \Pi}^{}_{7,23}(-q^{}_0) + i{\bf \Pi}^{}_{26,23}(-q^{}_0)  - i{\bf \Pi}^{}_{7,10}(-q^{}_0)\right\},
\end{eqnarray}
where the factor $1/4$ appears because of the global factor 2 in Eq.~(\ref{eq:Gauss2}) and factor $1/2$ from the Gauss integral, 
and the propagator is defined as inverse kernel matrix from Eq.~(\ref{eq:Gauss2})
\begin{equation}
{\bf \Pi}(P) = 
\left( 
\begin{array}{cc}
 {\bf A} & {\bf B} \\
 {\bf C} & {\bf D}
\end{array}
\right)^{-1}_P.
\end{equation} 
In the dc limit we get in terms of momentum independent masses
\begin{equation}
\bar\sigma^{\rm dc}_{12} = \frac{s}{4g^2}\frac{(M^{}_1+M^{}_5)^2}{(M^{}_1M^{}_5+\Delta^2_1)^2}. 
\end{equation}
Using Eqs.~(\ref{eq:MassAlg2}) this expression reduces to 
\begin{equation}
\bar\sigma^{\rm dc}_{12} = {\rm sgn}(m),
\end{equation}
which is the Hall conductivity of two separate cones.

\section{Conclusions}

\no
In this paper we investigate effects which arise if 2+1 dimensional Dirac or 
Weyl fermions are coupled to dispersionless in-plane phonons. The striking similarity of 
the electron-phonon coupling with relativistic gauge field theories makes it very appealing 
for deeper studies \cite{amorim16}. 
Intuitively, it is expected that sufficiently strong interactions can open up a gap in the spectrum 
of fermions and thus trigger the onset of an insulating phase, in fact, this scenario has been 
suggested in a number of previous work~\cite{Mudry2007,JakiwPi2007}. Furthermore, it is known that the gap 
in the spectrum of nodal fermions gives rise to topological, i.e. metric independent excitations 
in terms of gauge fields, which are usually referred to as Chern-Simons terms. For gapless 
fermions though, such excitations are usually linked to a quantum anomaly due to the 
a Pauli-Villars regularization of ultraviolet divergences, which requires introduction of 
some auxiliary heavy particles and violates {\it explicitly} the parity symmetry. 
For lattice models, which we investigate in our paper, there are no ultraviolet divergences 
because of the finite band width. Thus, there is no need to introduce any additional artificial 
degrees of freedom. In our approach, the quasiparticle mass and correspondingly the topological 
excitations appear due to {\it spontaneous} breaking of the parity symmetry. The question that 
remains is whether or not such excitations support quantized Hall currents. In the case of 
the single cone model there is a consensus on the relationship between Chern-Simons excitations 
and Hall currents~\cite{Semenoff1984,Haldane1988,Son2015,Fradkin2013,Halperin2017}. However, on lattices Dirac cones always 
appear in even numbers with different chiralities and therefore it depends on the form of 
the mass term, whether or not a Hall current exists. In the binodal model, which we consider here,
the masses are created spontaneously with an opposite sign at the two nodes. This leads to several
Chern-Simons terms which appear for different components of an eight-dimensional matrix field,
in contrast to the usual Chern-Simons term of a three-component gauge field. Some of these
Chern-Simons terms create quantized Hall currents via the Kubo formula but not all.
Therefore, our approach combines in one picture the notions of the spontaneous breaking of the
parity symmetry, spontaneous mass generation, and the emergence of a quantized Hall conductivity. 
An open problem is the role of those Chern-Simons terms which do not contribute to the
Hall conductivity. They may be relevant for correlations that are not connected to transport
properties. This issue is beyond the scope of this work which focuses on the quantum Hall effect 
associated with pseudo gauge fields due to phonons.

\vspace{2mm}
\no
Another question concerns the experimental relevance of our results. It is generally accepted that the 
electron-phonon interaction in graphene is weak and cannot cause any considerable effect
~\cite{Ferrari2007,Hakonen2014}. Therefore, it is not a suitable test laboratory for the theory 
developed here. However, other materials with a nodal band structure, like monolayers of 
silicon (silicene)~\cite{Silicene1,Silicene2}, germanium (germanene)~\cite{Germanin}, or tin (stanene)~\cite{Stanene2013,Stanene2015}
have a stronger electron-phonon interaction, which may be candidates for testing our predictions.
A promising candidate is black phosphorus, which usually crystallize on puckered 
honeycomb lattice similar to graphene. The higher rigidity and lattice anisotropy of the black phosphorus structure 
suggests stronger phonon effects and correspondingly stronger electron-phonon coupling
~\cite{Dresselhaus2016,Ribeiro2018}. Although a direct semiconductor with a spectral gap of 
roughly 0.39eV under 
normal conditions, black phosphorus can become a nodal semiconductor if a uniaxial strain is 
applied to the layer~\cite{{Kim2015,Kim2017,Ehlen2018}}.

\section*{ACKNOWLEDGMENTS}

This work was supported by a grant of the Julian Schwinger Foundation for Physical Research.

\appendix

\section{Manipulations with the matrix element of the Kubo formula}
\label{app:Kubo}

\no
In order to bring the Kubo formula into the $\cal Q$-matrix representation we complement the current densities in the matrix element of the Kubo formula in accord with the $\cal Q$-field shift in Eq.~(\ref{eq:intmid1}):  
\begin{eqnarray}
\nn
\langle {\rm Tr}{\cal D}^{}_{x,\mu}{\rm Tr}{\cal D}^{}_{y,\nu} \rangle^{}_{\bar\psi\psi} &=& 
\frac{2}{g^2} \langle {\rm Tr}\left[{\cal Q}-\frac{g}{\sqrt{2}}{\cal D}-{\cal Q}\right]^{}_{x,\mu}{\rm Tr}\left[{\cal Q}-\frac{g}{\sqrt{2}}{\cal D}-{\cal Q}\right]^{}_{y,\nu} \rangle^{}_{{\cal Q},\bar\psi\psi} \\
\nn
&=& \frac{2}{g^2} \langle  {\rm Tr}{\cal Q}^{}_{x,\mu}  {\rm Tr}{\cal Q}^{}_{y,\nu}\rangle^{}_{{\cal Q},\bar\psi\psi}  +
\frac{2}{g^2} \langle {\rm Tr}\left[{\cal Q}-\frac{g}{\sqrt{2}}{\cal D}\right]^{}_{x,\mu}{\rm Tr}\left[{\cal Q}- \frac{g}{\sqrt{2}}{\cal D}\right]^{}_{y,\nu}
\rangle^{}_{{\cal Q},\bar\psi\psi}  \\
&+& \frac{2}{g^2} \langle {\rm Tr}\left[{\cal Q}-\frac{g}{\sqrt{2}}{\cal D}\right]^{}_{x,\mu} {\rm Tr}{\cal Q}^{}_{y,\nu}\rangle^{}_{{\cal Q},\bar\psi\psi}  
+ \frac{2}{g^2} \langle    {\rm Tr}{\cal Q}^{}_{x,\mu}   {\rm Tr}\left[{\cal Q}-\frac{g}{\sqrt{2}}{\cal D}\right]^{}_{y,\nu}\rangle^{}_{{\cal Q},\bar\psi\psi} .
\end{eqnarray}
We can see that all terms but the first are zero by functional integration: According to Eq.~(\ref{eq:intmid1}), since the integral measure in the functional integral over $\cal Q$ is invariant with respect to any shifts of variables, expression ${\cal Q}^{}_\mu-\frac{g}{\sqrt{2}}{\cal D}^{}_\mu$ can be regarded as an independent interaction variable. The functional integration must be carried out over the local (i.e. with no derivatives) and diagonal in position space Gaussian weight. The matrix element of the Kubo formula is not diagonal in the position space, since it describes the charge transport from the point $x$ to the point $y$ which need to be spatially and temporarily separated from each other. This means that the integration has to be performed over every ${\cal Q}-\frac{g}{\sqrt{2}}{\cal D}^{}$, which is zero by symmetry of the bosonic Gaussian integral. In the last expression, the integration of fermions can be carried out and we acquire 
\begin{equation}
\label{eq:KuboMtrElt1}
 \langle {\rm Tr}{\cal D}^{}_{x,\mu}{\rm Tr}{\cal D}^{}_{y,\nu} \rangle^{}_{{\cal Q},\bar\psi\psi} = \frac{2}{g^2} \langle  {\rm Tr}{\cal Q}^{}_{x,\mu}  {\rm Tr}{\cal Q}^{}_{y,\nu}\rangle^{}_{{\cal Q}}, 
\end{equation}
where the integration on the right hand side has to be performed over the non-linear action Eq.~(\ref{eq:LogAct1}). The traces in Eq.~(\ref{eq:KuboMtrElt1}) can be easily performed using the decomposition Eq.~(\ref{eq:Decomp}):
\begin{equation}
\label{eq:KuboMtrElt2}
\langle {\rm Tr}{\cal D}^{}_{x,\mu}{\rm Tr}{\cal D}^{}_{y,\nu} \rangle^{}_{{\cal Q},\bar\psi\psi} = \frac{8}{g^2} \langle{\cal Q}^{3}_{x,\mu}{\cal Q}^{3}_{y,\nu}\rangle^{}_{{\cal Q}}, 
\end{equation}
where ${\cal Q}^3$ is the field component which couples to the identity matrix $\gamma^{}_3$. Hence, in $\cal Q$-field representation the Kubo formula reads: 
\begin{equation}
\bar\sigma^{}_{\mu\nu} = \frac{16\pi}{\omega g^2} \int d^3x \left[e^{-i\omega(x^{}_0-y^{}_0)} -1\right]\langle{\cal Q}^{3}_{x,\mu}{\cal Q}^{3}_{y,\nu}\rangle^{}_{{\cal Q}}.
\end{equation}

\vspace{2mm}
\no
For binodal model, each current transforms from the fermionic into the $\cal Q$-matrix representation as follows:
\begin{equation}
\label{eq:app:current}
(\psi^\dag j^{}_\mu\psi) = -{\rm Tr}(j^{}_\mu\psi\psi^\dag) = -\frac{i}{2g}{\rm Tr}\left(j^{}_\mu\Pi^{}_a\left[{\cal Q}^{}_a - ig\Pi^{}_a\psi \psi^\dag - {\cal Q}^{}_a \right]\right) \rightarrow 
\frac{i}{2g}{\rm Tr}\left(j^{}_\mu\Pi^{}_a{\cal Q}^{}_a\right).
\end{equation}

\section{Inverse phonon propagator structure of the single cone model}
\label{app:sinmod}

\no
To quadratic order in gradient expansion each block of the inverse phonon propagator of the single node model  has following non-zero matrix elements (to simplify the notation we use here $s={\rm sgn}(m), m = |m|, P^2= p^2_0+p^2_1+p^2_2$) : 
\begin{equation}
\label{eq:BlockA}
{\bf A} = \left( 
\begin{array}{cccc}
 M^{}_1 + \frac{p_0^2+p_1^2}{48\pi m} &  0  &-\frac{sp^{}_1}{16\pi} + \frac{p^{}_0p^{}_2}{48\pi m} \hspace{2mm} & \frac{isp^{}_0}{16\pi} + \frac{i p^{}_1 p^{}_2}{48 \pi m} \\
 \\
 0 &   M^{}_2 + \frac{P^2}{48\pi m}    &   0   &   0 \\
 \\
 \frac{sp^{}_1}{16\pi} + \frac{p^{}_0p^{}_2}{48\pi m} &  0  &  M^{}_3 + \frac{p^2_1+p^2_2}{48\pi m} \hspace{2mm} & \frac{isp^{}_2}{16\pi} - \frac{ip^{}_0p^{}_1}{48\pi m} \\
 \\
- \frac{isp^{}_0}{16\pi} + \frac{i p^{}_1 p^{}_2}{48 \pi m} & 0 & -\frac{isp^{}_2}{16\pi} - \frac{ip^{}_0p^{}_1}{48\pi m}\hspace{2mm} &  M^{}_4 - \frac{p^2_0 + p^2_2}{48\pi m}
\end{array}
\right),
\end{equation}

\begin{equation}
\label{eq:BlockB}
{\bf B} = \left( 
\begin{array}{cccc}
  \frac{sp^{}_0}{16\pi m} + \frac{p^{}_1p^{}_2}{48\pi m} &  \frac{sp^{}_1}{16\pi} - \frac{p^{}_0 p^{}_2}{48\pi m}     &   0   &   i \Delta^{}_1 - i\frac{p^2_0 + p^2_1}{48 \pi m}   \\
\\  
 0   &   0    &    \Delta^{}_2 + \frac{P^2}{48\pi m}  & 0  \\
\\
\frac{sp^{}_2}{16\pi} - \frac{p^{}_0 p^{}_1}{48\pi m}   &   \Delta^{}_3 - \frac{p^2_1 + p^2_2}{48\pi m}  & 0 &  -\frac{isp^{}_1}{16\pi} - \frac{ip^{}_0p^{}_2}{48\pi m} \\
\\
-i \Delta^{}_1 +i \frac{p^2_0+p^2_2}{48\pi m} &  \frac{isp^{}_2}{16\pi} + \frac{ip^{}_0 p^{}_1}{48\pi m} & 0 & - \frac{sp^{}_0}{16\pi} + \frac{p^{}_1 p^{}_2}{48\pi m} \\
\end{array}
\right),
\end{equation}

\begin{equation}
\label{eq:BlockC}
{\bf C} = \left( 
\begin{array}{cccc}
-\frac{sp^{}_0}{16\pi} + \frac{p^{}_1p^{}_2}{48\pi m} &  0  &  - \frac{sp^{}_2}{16\pi} - \frac{p^{}_0p^{}_1}{48\pi m} \hspace{2mm} & -i\Delta^{}_1 + i\frac{p^2_0 + p^2_2}{48\pi m} \\
\\
-\frac{sp^{}_1}{16\pi} - \frac{p^{}_0p^{}_2}{48\pi m} & 0  & \Delta^{}_3 - \frac{p^2_1 + p^2_2}{48\pi m}  & -\frac{isp^{}_2}{16\pi} + \frac{i p^{}_0p^{}_1}{48\pi m} \\
\\
 0 & \Delta^{}_2 + \frac{P^2}{48\pi m}  &  0  & 0 \\
\\
i\Delta^{}_1-i\frac{p^2_0+p^2_1}{48\pi m} & 0 & \frac{isp^{}_1}{16\pi} - \frac{ip^{}_0p^{}_2}{48\pi m} & \frac{sp^{}_0}{16\pi} + \frac{p^{}_1p^{}_2}{48\pi m}
\end{array}
\right),
\end{equation}

\begin{equation}
\label{eq:BlockD}
{\bf D} = \left( 
\begin{array}{cccc}
 M^{}_1 + \frac{p^2_0+p^2_2}{48\pi m} \hspace{2mm} &  \frac{sp^{}_2}{16\pi} + \frac{p^{}_0 p^{}_1}{48\pi m}  &   0   &  \frac{isp^{}_0}{16\pi} - \frac{ip^{}_1 p^{}_2}{48\pi m} \\
\\
 -\frac{sp^{}_2}{16\pi} + \frac{p^{}_0 p^{}_1}{48\pi m} & M^{}_3 + \frac{p^2_1 + p^2_2}{48\pi m} & 0 & \frac{isp^{}_1}{16\pi} + \frac{ip^{}_0p^{}_2}{48\pi m} \\
\\
 0 & 0 & M^{}_2 + \frac{P^2}{48\pi m} & 0 \\
\\
-\frac{isp^{}_0}{16\pi} - \frac{ip^{}_1 p^{}_2}{48\pi m} & -\frac{isp^{}_1}{16\pi} + \frac{ip^{}_0p^{}_2}{48\pi m} & 0 & M^{}_4 - \frac{p^2_0 + p^2_1}{48\pi m} 
\end{array}
\right),
\end{equation}
where the momentum independent quantities are defined as 
\begin{equation}
\label{eq:app:mas}
\begin{array}{lll}
\displaystyle M^{}_1 = \int\frac{d^3Q}{(2\pi)^3}~\frac{m^2+2q^2+q^2_0}{2[Q^2+m^2]^2},  & \displaystyle   M^{}_2 = \int\frac{d^3Q}{(2\pi)^3}~\frac{q^2+3m^2+q^2_0}{2[Q^2+m^2]^2}, & 
\displaystyle M^{}_3 = \int\frac{d^3Q}{(2\pi)^3}~\frac{q^2+m^2+3q^2_0}{2[Q^2+m^2]^2}, \\
\\
\displaystyle M^{}_4 = \int\frac{d^3Q}{(2\pi)^3}~\frac{2q^2+3m^2+3q^2_0}{2[Q^2+m^2]^2}, & \displaystyle \Delta^{}_1 = \int\frac{d^3Q}{(2\pi)^3}~\frac{m^2 + q^2_0}{2[Q^2+m^2]^2}, & 
\displaystyle \Delta^{}_2 = \int\frac{d^3Q}{(2\pi)^3}~\frac{m^2-q^2-q^2_0}{2[Q^2+m^2]^2}, \\ 
\\
\displaystyle \Delta^{}_3 = \int\frac{d^3Q}{(2\pi)^3}~\frac{q^2+m^2-q^2_0}{2[Q^2+m^2]^2}.
\end{array}
\end{equation}
Of all massive 8 phonon modes of the inverse propagator in Eq.~(\ref{eq:InvProp}), 6 are degenerated with the mass of the initial microscopic model
\begin{equation}
{\bf M}^{}_{1\cdots6} = \int\frac{d^3Q}{(2\pi)^3} ~\frac{1}{Q^2+m^2} = \frac{1}{2}\int\frac{d^2q}{(2\pi)^2}\frac{1}{\sqrt{q^2+m^2}} = \frac{1}{g}, 
\end{equation}
yet another heavy mode with the mass of the initial microscopic model as well
\begin{equation}
{\bf M}^{}_7 = \int\frac{d^3Q}{(2\pi)^3}~\frac{2q^2_0}{[Q^2+m^2]^2} = \frac{1}{2}\int\frac{d^2q}{(2\pi)^2}\frac{1}{\sqrt{q^2+m^2}} = \frac{1}{g},
\end{equation}
and one phonon mass which is proportional to the fermionic mass:
\begin{equation}
{\bf M}^{}_8 = \int\frac{d^3Q}{(2\pi)^3} ~\frac{2m^2}{[Q^2+m^2]^2} \sim \frac{|m|}{4\pi},
\end{equation}
where we neglected corrections of the order $\Lambda^{-1}$. All phonon masses are positive, hence there is no functional integral convergence issue.

\section{Saddle-point equations for the binodal model with the same chirality}
\label{app:SPEmonochir}

\no
The structure of the saddle-point equation Eq.~(\ref{eq:BiNSP}) does not change if one chooses to work within the representation with two fermions with the same chirality. In this case, as discussed above, the fermionic subset of the full set of $SU(4)$ group generators reads $\{\Sigma^{}_{01},\Sigma^{}_{02},\Sigma^{}_{03}\}$ and its commuting subset $\{\Sigma^{}_{20},\Sigma^{}_{30},\Sigma^{}_{10}\}$. The matrices by which phonons couple to fermions then become $\vec\Pi=\{\Sigma^{}_{11},\Sigma^{}_{12}\}$. This representation was used e.g. in Refs.~\cite{JakiwPi2007,Mudry2007,Ziegler2011,Ziegler2011a}. Formally, in terms of $\cal Q$-matrices, the saddle-point equation Eq.~(\ref{eq:SPE}) remains the same, with different matrices $\Pi^{}_\mu$ though. The order parameter of the Chern phase couples in this case to $\Sigma^{}_{03}$ matrix, i.e. for each channel we have
\begin{eqnarray}
 \alpha = 1: & \displaystyle 
 m^{}_1 \Sigma^{}_{11}\Sigma^{}_{12}  = ig\int\frac{d^3Q}{(2\pi)^3}~\left[G^{-1}_0 +i \left(m^{}_1 \Sigma^{}_{12}\Sigma^{}_{11} + m^{}_2 \Sigma^{}_{11} \Sigma^{}_{12}\right)\right]^{-1}, \\
 \alpha = 2: & \displaystyle 
 m^{}_2 \Sigma^{}_{12}\Sigma^{}_{11} = ig\int\frac{d^3Q}{(2\pi)^3}~\left[G^{-1}_0 +i \left(m^{}_1 \Sigma^{}_{12}\Sigma^{}_{11} + m^{}_2 \Sigma^{}_{11} \Sigma^{}_{12}\right)\right]^{-1}.
\end{eqnarray}
which reduces to 
\begin{eqnarray}
m^{}_1=-m^{}_2 &=& 0.5m \\
(m^{}_1-m^{}_2) \Sigma^{}_{03} &=& 2g \int\frac{d^3Q}{(2\pi)^3}~\left[G^{-1}_0 + \left(m^{}_1 - m^{}_2 \right)\Sigma^{}_{03} \right]^{-1},
\end{eqnarray}
since $\Sigma^{}_{03} $ anticommutes with the fermionic part of the Hamiltonian, the inversion of the Green's function is trivial and we arrive at Eq.~(\ref{eq:BiNSP}). The structure of Gaussian fluctuation term will change accordingly, but the ultimate result for the Hall conductivity will remain the same.

\vspace{2mm}
\no
One might wonder whether the fermionic mass which couples to the matrix $\Sigma^{}_{03}$ can be spontaneously generated in the representation which we use in the main part. Besides the fact that because of the Pauli algebra ${\rm Tr}[\Sigma^{}_{01}\Sigma^{}_{32}\Sigma^{}_{03}]=0$ such mass is not capable of generating a Chern-Simons term, it is easy to show hat this regime would only have tachyon-like solutions
\begin{equation}
|\tilde m| = -\int\frac{d^3Q}{(2\pi)^3}~\frac{2g|\tilde m|}{Q^2-|\tilde m|^2},
\end{equation}
which does not yield any positive or well defined value for the parameter $\tilde m$.

\section{Inverse phonon propagator structure of the binodal model}
\label{app:binmod}

\no
To quadratic order in gradient expansion each block of the inverse phonon propagator of the binodal model has following non-zero matrix elements (here again $s={\rm sgn}(m),\;\; m=|m|, \;\; P^2=p^2_0+p^2_1+p^2_2$): 
\begin{itemize}
\item Block ${\bf A}$ :
\begin{equation}
\begin{array}{llll}
 A^{}_{1,1} = M^{}_1 + \frac{p^2_0+p^2_1}{24\pi m},   &    A^{}_{1,15} = \frac{isp^{}_1}{8\pi} -\frac{ip^{}_0 p^{}_2}{24\pi m}, &  
 A^{}_{1,16} = -\frac{isp^{}_0}{8\pi} - \frac{ip^{}_1p^{}_2}{24\pi m}, & A^{}_{2,2} = M^{}_2 + \frac{P^2}{24\pi m},\\
 \\
 A^{}_{3,3} = M^{}_3 - \frac{p^2_1+p^2_2}{24\pi m}, &  A^{}_{3,4} = -\frac{sp^{}_2}{8\pi} + \frac{p^{}_0 p^{}_1}{24\pi m}, &
 A^{}_{3,13} = \frac{isp^{}_1}{8\pi} + \frac{ip^{}_0 p^{}_2}{24\pi m}, &  A^{}_{4,3} = \frac{sp^{}_2}{8\pi} + \frac{p^{}_0 p^{}_1}{24\pi m}, \\
 \\
 A^{}_{4,4} = M^{}_1 - \frac{p^2_0+p^2_2}{24\pi m}, &  A^{}_{4,13} = -\frac{isp^{}_0}{8\pi} + \frac{ip^{}_1p^{}_2}{24\pi m}, & 
 A^{}_{5,5} = M^{}_4 - \frac{P^2}{24\pi m}, &  A^{}_{6,6} = M^{}_5 + \frac{p^2_0+p^2_1}{24\pi m}, \\
 \\
 A^{}_{6,11} = -\frac{isp^{}_0}{8\pi}-\frac{ip^{}_1p^{}_2}{24\pi m}, & A^{}_{6,12} = -\frac{isp^{}_1}{8\pi}-\frac{ip^{}_0p^{}_2}{24\pi m}, & 
 A^{}_{7,7} = M^{}_5 + \frac{p^2_0 + p^2_2}{24\pi m}, &  A^{}_{7,8} = \frac{sp^{}_2}{8\pi} + \frac{p^{}_0p^{}_1}{24\pi m}, \\
 \\
 A^{}_{7,10} = \frac{isp^{}_0}{8\pi} - \frac{ip^{}_1p^{}_2}{24\pi m}, &  A^{}_{8,7} = -\frac{sp^{}_2}{8\pi}+\frac{p^{}_0p^{}_1}{24\pi m}, &
 A^{}_{8,8} = M^{}_6+\frac{p^2_1+p^2_2}{24\pi m}, & A^{}_{8,10} = \frac{isp^{}_1}{8\pi}+\frac{ip^{}_0p^{}_2}{24\pi m}, \\ 
 \\
  A^{}_{9,9} = M^{}_2 + \frac{P^2}{24\pi m}, &  A^{}_{10,7} = -\frac{isp^{}_0}{8\pi}-\frac{ip^{}_1p^{}_2}{24\pi m}, & 
  A^{}_{10,8} = -\frac{isp^{}_1}{8\pi} + \frac{ip^{}_0p^{}_2}{24\pi m}, &  A^{}_{10,10}=M^{}_1-\frac{p^2_0 + p^2_1}{24\pi m}, \\
  \\
  A^{}_{11,6} = \frac{isp^{}_0}{8\pi}-\frac{ip^{}_1p^{}_2}{24\pi m}, &  A^{}_{11,11}=M^{}_1 - \frac{p^2_0+p^2_2}{24\pi m}, &
  A^{}_{11,12} = -\frac{sp^{}_2}{8\pi} - \frac{p^{}_0p^{}_1}{24\pi m}, &  A^{}_{12,6} = \frac{isp^{}_1}{8\pi}+\frac{ip^{}_0p^{}_2}{24\pi m}, \\
  \\
  A^{}_{12,11} = \frac{sp^{}_2}{8\pi}-\frac{p^{}_0p^{}_1}{24\pi m}, &  A^{}_{12,12}=M^{}_3-\frac{p^2_1+p^2_2}{24\pi m}, & 
  A^{}_{13,3}=-\frac{isp^{}_1}{8\pi}+\frac{ip^{}_0p^{}_2}{24\pi m}, &  A^{}_{13,4} = \frac{isp^{}_0}{8\pi}+\frac{ip^{}_1p^{}_2}{24\pi m}, \\
  \\ 
  A^{}_{13,13} = M^{}_5 + \frac{p^2_0+p^2_1}{24\pi m}, &  A^{}_{14,14}=M^{}_4 - \frac{P^2}{24\pi m}, & 
  A^{}_{15,1}=-\frac{isp^{}_1}{8\pi}-\frac{ip^{}_0p^{}_2}{24\pi m}, &  A^{}_{15,15}=M^{}_6+\frac{p^2_1+p^2_2}{24\pi m}, \\
  \\
  A^{}_{15,16} = \frac{sp^{}_2}{8\pi}-\frac{p^{}_0p^{}_1}{24\pi m}, &  A^{}_{16,1}=\frac{isp^{}_0}{8\pi}-\frac{ip^{}_1p^{}_2}{24\pi m}, & 
  A^{}_{16,15}=-\frac{sp^{}_2}{8\pi}-\frac{p^{}_0p^{}_1}{24\pi m}, &  A^{}_{16,16} = M^{}_5+\frac{p^2_0+p^2_2}{24\pi m}.
 \end{array}
\end{equation}

\item Block ${\bf B}$ :
\begin{equation}
\begin{array}{llll}
 B^{}_{1,1} = -\frac{sp^{}_0}{8\pi} -\frac{p^{}_1p^{}_2}{24\pi m}, &  B^{}_{1,2} = -\frac{isp^{}_1}{8\pi} + \frac{ip^{}_0p^{}_2}{24\pi m}, & 
 B^{}_{1,16} = -i\Delta^{}_1 +i \frac{p^2_0+p^2_1}{24\pi m}, &  B^{}_{2,15} = -\Delta^{}_2 + \frac{P^2}{24\pi m}, \\
 \\
 B^{}_{3,4} = \frac{sp^{}_1}{8\pi} + \frac{p^{}_0p^{}_2}{24\pi m}, & B^{}_{3,13} = \frac{isp^{}_2}{8\pi} -\frac{ip^{}_0p^{}_1}{24\pi m}, & 
 B^{}_{3,30} = \Delta^{}_3 + \frac{p^2_1 + p^2_2}{24\pi m}, &  B^{}_{4,4} = -\frac{sp^{}_0}{8\pi} + \frac{p^{}_1 p^{}_2}{24\pi m}, \\
 \\
 B^{}_{4,13} = -i\Delta^{}_1 + i\frac{p^2_0 + p^2_2}{24\pi m}, &  B^{}_{4,14} = -\frac{sp^{}_2}{8\pi} - \frac{p^{}_0p^{}_1}{24\pi m}, & 
 B^{}_{5,12} = \Delta^{}_2 - \frac{P^2}{24\pi m}, & B^{}_{6,5} = \frac{isp^{}_1}{8\pi} - \frac{ip^{}_0p^{}_2}{24\pi m}, \\
 \\
 B^{}_{6,6} = \frac{sp^{}_0}{8\pi} + \frac{p^{}_1p^{}_2}{24\pi m}, & B^{}_{6,11} = -i\Delta^{}_{1} + i\frac{p^2_0 + p^2_1}{24\pi m}, &
 B^{}_{7,7} = \frac{sp^{}_0}{8\pi}-\frac{p^{}_1 p^{}_2}{24\pi m}, &  B^{}_{7,9} = \frac{sp^{}_2}{8\pi} + \frac{p^{}_0p^{}_1}{24\pi m}, \\
 \\
 B^{}_{7,10} = i \Delta^{}_1 - i\frac{p^2_0 + p^2_2}{24\pi m}, &  B^{}_{8,7} = \frac{sp^{}_1}{8\pi} + \frac{p^{}_0p^{}_2}{24\pi m}, &
 B^{}_{8,9} = \Delta^{}_3 + \frac{p^2_1+p^2_2}{24\pi m}, & B^{}_{8,10} = \frac{isp^{}_2}{8\pi} - \frac{ip^{}_0p^{}_1}{24\pi m}, \\
 \\
 B^{}_{9,8} = \Delta^{}_2 - \frac{P^2}{24\pi m}, &  B^{}_{10,7} = -i\Delta^{}_1 + i\frac{p^2_0+p^2_1}{24\pi m}, & 
 B^{}_{10,9} = -\frac{isp^{}_1}{8\pi}+\frac{ip^{}_0p^{}_2}{24\pi m}, & B^{}_{10,10}=-\frac{sp^{}_0}{8\pi}-\frac{p^{}_1p^{}_2}{24\pi m}\\ 
 \\
 B^{}_{11,5} = \frac{sp^{}_2}{8\pi}+\frac{p^{}_0p^{}_1}{24\pi m}, & B^{}_{11,6} = i\Delta^{}_1-i\frac{p^2_0+p^2_2}{24\pi m}, & 
 B^{}_{11,11} = -\frac{sp^{}_0}{8\pi} + \frac{p^{}_1p^{}_2}{24\pi m}, &  B^{}_{12,5}=\Delta^{}_3+\frac{p^2_1+p^2_2}{24\pi m}, \\
 \\
 B^{}_{12,6}=\frac{isp^{}_2}{8\pi}-\frac{ip^{}_0p^{}_1}{24\pi m}, & B^{}_{12,11}=-\frac{sp^{}_1}{8\pi}-\frac{p^{}_0p^{}_2}{24\pi m}, &
 B^{}_{13,4}=i\Delta^{}_1-i\frac{p^2_0+p^2_1}{24\pi m}, & B^{}_{13,13} = \frac{sp^{}_0}{8\pi}+\frac{p^{}_1p^{}_2}{24\pi m}, \\
 \\
 B^{}_{13,14}=\frac{isp^{}_1}{8\pi}-\frac{ip^{}_0p^{}_2}{24\pi m}, &  B^{}_{14,3} = \Delta^{}_2 -\frac{P^2}{24\pi m}, & 
 B^{}_{15,1} = -\frac{isp^{}_2}{8\pi}+\frac{ip^{}_0p^{}_1}{24\pi m}, & B^{}_{15,2}=-\Delta^{}_3-\frac{p^2_1+p^2_2}{24\pi m}, \\
  \\
 B^{}_{15,16}=-\frac{sp^{}_1}{8\pi}-\frac{p^{}_0p^{}_2}{24\pi m}, & B^{}_{16,1}=i\Delta^{}_1-i\frac{p^2_0+p^2_2}{24\pi m}, & 
 B^{}_{16,2} = \frac{sp^{}_2}{8\pi}+\frac{p^{}_0p^{}_1}{24\pi m}, & B^{}_{16,16} = \frac{sp^{}_0}{8\pi}-\frac{p^{}_1p^{}_2}{24\pi m}.
 \end{array}
\end{equation}

\item Block ${\bf C}$ :
\begin{equation}
\begin{array}{llll}
 C^{}_{1,1} = \frac{sp^{}_0}{8\pi}-\frac{p^{}_1p^{}_2}{24\pi m}, &  C^{}_{1,15}=\frac{isp^{}_2}{8\pi}+\frac{ip^{}_0p^{}_1}{24\pi m}, & 
 C^{}_{1,16} = i\Delta^{}_1-i\frac{p^2_0+p^2_2}{24\pi m}, &  C^{}_{2,1} = \frac{isp^{}_1}{8\pi}+\frac{ip^{}_0p^{}_2}{24\pi m}, \\
 \\ 
 C^{}_{2,15}=-\Delta^{}_3-\frac{p_1^2+p_2^2}{24\pi m}, & C^{}_{2,16}=-\frac{sp^{}_2}{8\pi}+\frac{p^{}_0p^{}_1}{24\pi m}, &
 C^{}_{3,14} = \Delta^{}_2-\frac{P^2}{24\pi m}, & C^{}_{4,3}=-\frac{sp^{}_1}{8\pi}+\frac{p^{}_0p^{}_2}{24\pi m}, \\
 \\
 C{}_{4,4}=\frac{sp^{}_0}{8\pi}+\frac{p^{}_1p^{}_2}{24\pi m}, &  C^{}_{4,13}=i\Delta^{}_1-i\frac{p^2_0+p^2_1}{24\pi m}, & 
 C^{}_{5,6} = -\frac{isp^{}_1}{8\pi}-\frac{ip^{}_0p^{}_2}{24\pi m}, & C^{}_{5,11} = -\frac{sp^{}_2}{8\pi} + \frac{p^{}_0p^{}_1}{24\pi m},\\  
 \\
 C^{}_{5,12}=\Delta^{}_3+\frac{p^2_1+p^2_2}{24\pi m}, & C^{}_{6,6}=-\frac{sp^{}_0}{8\pi}+\frac{p^{}_1p^{}_2}{24\pi m}, & 
 C^{}_{6,11} = i\Delta^{}_1 -i\frac{p^2_0+p^2_2}{24\pi m}, &  C^{}_{6,12} = -\frac{isp^{}_2}{8\pi}-\frac{ip^{}_0p^{}_1}{24\pi m}, \\
 \\
 C^{}_{7,7}=-\frac{sp^{}_0}{8\pi}-\frac{p^{}_1p^{}_2}{24\pi m}, & C^{}_{7,8} = -\frac{sp^{}_1}{8\pi}+\frac{p^{}_0p^{}_2}{24\pi m}, &
 C^{}_{7,10}=-i\Delta^{}_1+i\frac{p^2_0+p^2_1}{24\pi m}, & C^{}_{8,9}=\Delta^{}_2-\frac{P^2}{24\pi m}, \\
 \\
 C^{}_{9,7}=-\frac{sp^{}_2}{8\pi}+\frac{p^{}_0p^{}_1}{24\pi m}, & C^{}_{9,8}=\Delta^{}_3+\frac{p^2_1+p^2_2}{24\pi m}, & 
 C^{}_{9,10}=\frac{isp^{}_1}{8\pi}+\frac{ip^{}_0p^{}_2}{24\pi m}, & C^{}_{10,7}=i\Delta^{}_1-i\frac{p^2_0+p^2_2}{24\pi m}, \\
  \\
 C^{}_{10,8}=-\frac{isp^{}_2}{8\pi}-\frac{ip^{}_0p^{}_1}{24\pi m}, & C^{}_{10,10}=\frac{sp^{}_0}{8\pi}-\frac{p^{}_1p^{}_2}{24\pi m}, & 
 C^{}_{11,6}=-i\Delta^{}_1+i\frac{p^2_0+p^2_1}{24\pi m}, &  C^{}_{11,11} = \frac{sp^{}_0}{8\pi}+\frac{p^{}_1p^{}_2}{24\pi m},  \\
 \\
 C^{}_{11,12}=\frac{sp^{}_1}{8\pi}-\frac{p^{}_0p^{}_2}{24\pi m}, &  C^{}_{12,5} = \Delta^{}_2-\frac{P^2}{24\pi m}, &
 C^{}_{13,3}=-\frac{isp^{}_2}{8\pi}-\frac{ip^{}_0p^{}_1}{24\pi m}, & C^{}_{13,4}=-i\Delta^{}_1+i\frac{p^2_0+p^2_2}{24\pi m}, \\
 \\
 C^{}_{13,13}=-\frac{sp^{}_0}{8\pi}+\frac{p^{}_1p^{}_2}{24\pi m}, &  C^{}_{14,3}=\Delta^{}_3+\frac{p^2_1+p^2_1}{24\pi m}, &
 C^{}_{14,4}=\frac{sp^{}_2}{8\pi}-\frac{p^{}_0p^{}_1}{24\pi m}, & C^{}_{14,13}=-\frac{isp^{}_1}{8\pi}-\frac{p^{}_0p^{}_2}{24\pi m}, \\
 \\
 C^{}_{15,2}=-\Delta^{}_2+\frac{P^2}{24\pi m}, & C^{}_{16,1}=-i\Delta^{}_1+i\frac{p^2_0+p^2_1}{24\pi m}, & 
 C^{}_{16,15}=\frac{sp^{}_1}{8\pi}-\frac{p^{}_0p^{}_2}{24\pi m}, & C^{}_{16,16}=-\frac{sp^{}_0}{8\pi}-\frac{p^{}_1p^{}_2}{24\pi m}.
\end{array}
\end{equation}

\item Block ${\bf D}$ :
\begin{equation}
\begin{array}{llll}
 D^{}_{1,1} = M^{}_1-\frac{p^2_0+p^2_2}{24\pi m}, &  D^{}_{1,2} = -\frac{isp^{}_2}{8\pi}-\frac{ip^{}_0p^{}_1}{24\pi m}, & 
 D^{}_{1,16}=-\frac{isp^{}_0}{8\pi}+\frac{ip^{}_1p^{}_2}{24\pi m}, &  D^{}_{2,1} = \frac{isp^{}_2}{8\pi}-\frac{ip^{}_0p^{}_1}{24\pi m}, \\
 \\
 D^{}_{2,2} = M^{}_6+\frac{p^2_1+p^2_2}{24\pi m}, & D^{}_{2,16}=\frac{sp^{}_1}{8\pi}+\frac{p^{}_0p^{}_2}{24\pi m}, &  
 D^{}_{3,3} = M^{}_4-\frac{P^2}{24\pi m}, &  D^{}_{4,4} = M^{}_1 - \frac{p^2_0+p^2_1}{24\pi m}, \\
 \\
 D^{}_{4,13}=-\frac{ip^{}_0}{8\pi}-\frac{p^{}_1p^{}_2}{24\pi m}, &  D^{}_{4,14}=\frac{sp^{}_1}{8\pi}-\frac{p^{}_0}{24\pi m}, & 
 D^{}_{5,5} = M^{}_3-\frac{p^2_1+p^2_2}{24\pi m}, & D^{}_{5,6} = -\frac{isp^{}_2}{8\pi}+\frac{ip^{}_0p^{}_1}{24\pi m}, \\
 \\
 D^{}_{5,11}=\frac{sp^{}_1}{8\pi}+\frac{p^{}_0p^{}_2}{24\pi m}, & D^{}_{6,5}=\frac{isp^{}_2}{8\pi}+\frac{ip^{}_0p^{}_1}{24\pi m}, & 
 D^{}_{6,6} = M^{}_5+\frac{p^2_0+p^2_2}{24\pi m}, &  D^{}_{6,11}=-\frac{isp^{}_0}{8\pi}+\frac{ip^{}_1p^{}_2}{24\pi m}, \\
 \\
 D^{}_{7,7}=M^{}_5+\frac{p^2_0+p^2_1}{24\pi m}, & D^{}_{7,9}=-\frac{sp^{}_1}{8\pi}+\frac{p^{}_0p^{}_2}{24\pi m}, &
 D^{}_{7,10}=\frac{isp^{}_0}{8\pi}+\frac{ip^{}_1p^{}_2}{24\pi m}, & D^{}_{8,8}=M^{}_2+\frac{P^2}{24\pi m}, \\ 
 \\
 D^{}_{9,7}=\frac{sp^{}_1}{8\pi}+\frac{p^{}_0p^{}_2}{24\pi m}, &  D^{}_{9,9}=M^{}_6+\frac{p^2_1+p^2_2}{24\pi m}, & 
 D^{}_{9,10}=\frac{isp^{}_2}{8\pi}-\frac{ip^{}_0p^{}_1}{24\pi m}, & D^{}_{10,7}=-\frac{isp^{}_0}{8\pi}+\frac{ip^{}_1p^{}_2}{24\pi m}, \\
 \\
 D^{}_{10,9}=-\frac{isp^{}_2}{8\pi}-\frac{ip^{}_0p^{}_1}{24\pi m}, & D^{}_{10,10}=M^{}_1-\frac{p^2_0+p^2_2}{24\pi m}, & 
 D^{}_{11,5}=-\frac{sp^{}_1}{8\pi}+\frac{p^{}_0p^{}_2}{24\pi m}, &  D^{}_{11,6}=\frac{isp^{}_0}{8\pi}+\frac{ip^{}_1p^{}_2}{24\pi m}, \\
 \\
 D^{}_{11,11} = M^{}_1 - \frac{p^2_0+p^2_1}{24\pi m}, & D^{}_{12,12}=M^{}_4-\frac{P^2}{24\pi m}, &
 D^{}_{13,4}=\frac{isp^{}_0}{4\pi}-\frac{ip^{}_1p^{}_2}{24\pi m}, & D^{}_{13,13}=M^{}_5+\frac{p^2_0+p^2_2}{24\pi m}, \\
 \\
 D^{}_{13,14}=\frac{isp^{}_2}{8\pi}+\frac{ip^{}_0p^{}_1}{24\pi m}, &  D^{}_{14,4} = -\frac{sp^{}_1}{8\pi}-\frac{p^{}_0 p^{}_2}{24\pi m}, & 
 D^{}_{14,13}=-\frac{isp^{}_2}{8\pi}+\frac{ip^{}_0p^{}_1}{24\pi m}, & D^{}_{14,14}=M^{}_3-\frac{p^2_1+p^2_2}{24\pi m}, \\
 \\
 D^{}_{15,15}=M^{}_2+\frac{P^2}{24\pi m}, & D^{}_{16,1}=\frac{isp^{}_0}{8\pi}+\frac{ip^{}_1p^{}_2}{24\pi m}, & 
 D^{}_{16,2}=-\frac{sp^{}_1}{8\pi}+\frac{p^{}_0p^{}_2}{24\pi m}, &  D^{}_{16,16}=M^{}_5+\frac{p^2_0+p^2_1}{24\pi m}.
\end{array}
\end{equation}
\end{itemize}
Here we use a different notation for the mass terms:

\begin{equation}
\label{eq:app:2cmas}
\begin{array}{lll}
\displaystyle M^{}_1 = \int\frac{d^3Q}{(2\pi)^3}~\frac{2q^2+3q^2_0+3m^3}{[Q^2+m^2]^2}, \hspace{2mm} & \displaystyle M^{}_2 = \int\frac{d^3Q}{(2\pi)^3}~\frac{q^2+q^2_0+3m^2}{[Q^2+m^2]^2}, & 
\displaystyle M^{}_3 = \int\frac{d^3Q}{(2\pi)^3}~\frac{3q^2+q^2_0+3m^2}{[Q^2+m^2]^2}, \\
\\
\displaystyle M^{}_4 = \int\frac{d^3Q}{(2\pi)^3}~\frac{3q^2+3q^2_0+m^2}{[Q^2+m^2]^2}, \hspace{2mm} & \displaystyle M^{}_5 = \int\frac{d^3Q}{(2\pi)^3}~\frac{2q^2+q^2_0+m^2}{[Q^2+m^2]^2}, & 
\displaystyle M^{}_6 = \int\frac{d^3Q}{(2\pi)^3}~\frac{q^2+3q^2_0+m^2}{[Q^2+m^2]^2}, \\
\\
\displaystyle \Delta^{}_1 = \int\frac{d^3Q}{(2\pi)^3}~\frac{q^2_0+m^2}{[Q^2+m^2]^2}, \hspace{2mm} & \displaystyle \Delta^{}_2 = \int\frac{d^3Q}{(2\pi)^3}~\frac{q^2_0+q^2-m^2}{[Q^2+m^2]^2}, & 
\displaystyle \Delta^{}_3 = \int\frac{d^3Q}{(2\pi)^3}~\frac{q^2_0-q^2-m^2}{[Q^2+m^2]^2}.
\end{array}
\end{equation}


\end{document}